\documentclass[showpacs,twocolumn,amsmath,superscriptaddress,amssymb,amsfonts,
prb,floatfix]{revtex4-1}
\pdfoutput=1
\usepackage{graphicx, color,rotating} 
\usepackage{dcolumn} 
\usepackage{bm} 
\usepackage{epsfig}
\usepackage{bm}
\usepackage{dsfont}
\usepackage{relsize}
\usepackage{soul}
\usepackage{subfigure}
\usepackage[plainpages=false, pdfpagelabels, bookmarksopen, pdfborder={0 0 0},
linkcolor=blue, linkbordercolor={0 0 1}]{hyperref}

\usepackage{graphicx}
\usepackage{xcolor}
\usepackage{colortbl}
\usepackage{tikz}
\usetikzlibrary{arrows,decorations.pathmorphing,decorations.markings,backgrounds
,positioning,fit,trees}
\tikzset{->-/.style={decoration={
  markings,
  mark=at position #1 with {\arrow{>}}},postaction={decorate}}}
\tikzset{
   vertex/.style={circle, inner sep=0pt, minimum size=5pt,fill=black,label=#1},
vertex/.default=\text{},
   crossing/.style={circle, inner sep=0, minimum size=0,label=#1},
crossing/.default=\text{},
   named/.style={draw,circle, inner sep=0pt, minimum size=12pt},
   namedE/.style={draw,circle, inner sep=0pt, minimum size=7pt},
   bath/.style={draw,thick,->-=.5}, bath/.default={},
   bathr/.style={draw,thick,dashed,->-=.5},
   bathNA/.style={draw,thick}, bath/.default={},
   bathrNA/.style={draw,thick,dashed},
   time/.style={draw,dashed,thin},
   timeAxes/.style={draw,thick,->-=1}, timeAxes/.default={},
   system/.style={draw,thick}, system/.default={},
   syslabel/.style={midway,auto,green!40!black}}
\tikzset{every picture/.append style={node distance=1.5*1cm and 1.5*2cm,bend
angle=60}}

\newcommand{\bra}[1]{\langle #1|}
\newcommand{\ket}[1]{|#1\rangle}

\begin{document}
\title{Qubit dephasing due to Quasiparticle Tunneling}

\author{Sebastian Zanker}
\affiliation{Institut f\"ur Theoretische Festk\"orperphysik, Karlsruhe Institute of Technology, D-76128 Karlsruhe,
Germany}
\author{Michael Marthaler}
\affiliation{Institut f\"ur Theoretische Festk\"orperphysik, Karlsruhe Institute of Technology, D-76128 Karlsruhe,
Germany}


\date{\today}

\begin{abstract}
We study dephasing of a superconducting qubit due to quasiparticle tunneling
through a Josephson junction. While qubit decay due to tunneling processes is
well understood within a golden rule approximation, pure dephasing due to BCS quasiparticles
gives rise to a divergent golden rule rate. We calculate qubit dephasing due to quasiparticle tunneling beyond lowest order approximation in coupling between qubit  and quasiparticles.
Summing up a certain class of diagrams we show that qubit
dephasing due to purely longitudinal coupling to quasiparticles 
leads to a dephasing $\sim \exp(-x(t))$ where $x(t)$ is not linear in time on short time scales while it tends towards a selfconsistent calculated dephasing rate for longer times.
\end{abstract}

\maketitle

\section{Introduction}
Superconducting quantum circuits based on the Josephson effect are promising candidates for the realization of large scale quantum computers\cite{Science-2013-Devoret-1169-74-1}. While early qubit designs such as the charge\cite{1402-4896_1998_T76_024}, flux\cite{Science-1999-Mooij-1036-9} and phase qubit\cite{PhysRevLett.89.117901} were relatively sensitive to environmental charge and phase fluctuations,  qubits such as the transmon\cite{PhysRevA.76.042319} or fluxonium have overcome these issues. With the 3D implementation of the transmon,
decoherence times up to almost 100$\mu$s have been
demonstrated\cite{PhysRevLett.107.240501,PhysRevB.86.100506-1,
PhysRevB.86.180504-1} closing in on quantum error correction thresholds.
Even 2D implementations of transmon qubits at the threshold of quantum error correction have been demonstrated\cite{PhysRevLett.111.080502}.
With these decoherence times superconducting qubits have reached regimes where previously unobservable decoherence channels such as quasiparticle tunneling could become a relevant source of dephasing. Quasiparticles as a source of decoherence have been confirmed by several experiments that clearly demonstrate the influence of non-equilibrium quasiparticles on qubit energy relaxation either with temperature dependent measurements\cite{PhysRevLett.103.097002} or with quasiparticles injected on purpose \cite{1.3638063,PhysRevB.84.024501,PhysRevLett.103.097002,PhysRevB.78.024503}.
Quasiparticles as an intrinsic feature of superconducting devices are of particular interest because they could provide an ultimate limit to qubit coherence times. Besides equilibrium quasiparticles which, at usual qubit operation temperature, are exponentially suppressed and contribute little to the overall quasiparticle density, there always exist non equilibrium quasiparticles close to the BCS gap. Qubit decay and frequency shifts due to those non equilibrium quasiparticles have been studied in several theoretical papers
\cite{PhysRevB.72.014517,PhysRevB.85.144503,PhysRevB.84.064517} with golden rule
calculations. Diagonal elements of a qubit's density matrix decay to their
equilibrium values with a relaxation rate that is proportional to the
quasiparticle spectral density $S_{qp}(\omega)$ evaluated at the qubit frequency
$\epsilon_0$ with decay times ranging from several micro seconds up to
milliseconds depending on the qubit type\cite{PhysRevB.85.144503}. In addition
to energy relaxation with decay rate $\Gamma_1$ quasiparticle tunneling induces
pure dephasing. The decay of off-diagonal elements of the density matrix takes the form $\rho_{10}\sim e^{-\Gamma_1 t/2}e^{-x(t)}$ with the 
dephasing function $x(t)$ which in general is not linear in time. Nonetheless
for noise with a regular spectral density at low frequencies and for long times 
one can define a pure dephasing rate $\Gamma_{2^*}$ and the dephasing function takes the form $x(t)=\Gamma_{2^*}t$. 
Contrary to relaxation pure dephasing is proportional to the spectral density at low frequencies and a golden rule calculation yields $\Gamma_{2^*}\sim S(0)$.
Unfortunately, the BCS density of states leads to a spectral density that diverges logarithmically as $\omega$ tends to zero. As is known from 1/f noise, dephasing due to noise
with a divergent spectral density at low frequencies produces non-linear
exponential decay\cite{PhysRevB.72.134519}. For 1/f noise the dephasing function is quadratic, $x(t)\sim bt^2$.
Since the irregularity of the quasiparticle spectral density is only logarithmic
we expect a time dependence of dephasing due to quasiparticles somewhere between
the linear golden rule and the quadratic 1/f-noise result,  $x(t)\sim
b t^\alpha$ with $1\leq\alpha\leq2$.
%
Single quasiparticle tunneling changes the parity of the qubit state and recent
experimental\cite{ncomms2936-1} and theoretical\cite{1401.5575} works suggest
that this parity change can be important for qubit decoherence even for the
transmon with large ratio between Josephson and charge energy. We neglect these
effects in this work which is valid for small energy splittings between physical
states for the same qubit level but with different parity\cite{1401.5575}.

In this paper we us two different approaches to estimate pure dephasing due to
quasiparticle tunneling. First we use a real time diagrammatic
technique to find a selfconsistent pure dephasing rate.The result
is similar to the selfconsistent rate defined by Catelani\cite{1207.7084}. This
approach leads to a linear exponential decay but, as we have mentioned before,
dephasing due to a divergent spectral density usually is non-exponential at short times. To
calculate the non-linear behavior we sum up a certain group of diagrams for quasiparticle tunneling. 
With this summation we recover the results obtained for a bosonic bath coupled longitudinal to the qubit. Hence we find that we can describe
dephasing due to quasiparticles with relations already established for the
treatment of bosonic noise\cite{PhysRevB.72.134519}.

\section{The Model}\label{sec:model}
The effective Hamiltonian of a superconducting qubit coupled to quasiparticle
degrees of freedom can be split into three parts: 
\begin{equation}\label{eq:model-Heff}
 H = H_{S}+H_{R}+H_T.
\end{equation}
Here $H_{S}$ is the effective qubit Hamiltonian that includes all coherent many-body degrees of freedom that
contribute to the effective two level system. $H_{R}$
describes
the free quasiparticles in the superconducting leads of the system. The last
term $H_T$ describes quasiparticle tunneling across the Josephson junction and
couples qubit and quasiparticle
degrees of freedom. This coupling induces decay and dephasing in
the time evolution of the qubit. In general we can distinguish
two different kinds of tunneling processes: tunneling processes with energy
transfer inducing transitions between qubit states and elastic processes
that contribute to pure dephasing only. We neglect the change in parity due to
single quasiparticle tunneling processes. For a transmon qubit, whose
eigenstates are superpositions of states with even and odd parity, this
treatment is valid for small energy splitting between states with
different parity but belonging to the same qubit eigenstate\cite{1401.5575}.

The distinction between qubit, free quasiparticles and tunneling may, at first
glance, seem artificial as the Josephson junction, described with the same
quasiparticle tunnel Hamiltonian $H_T$, is part of the qubit. Indeed one has to
be careful to avoid double counting of tunneling processes. We will come back to
this issue in more detail in section (\ref{sec:qubit}) and (\ref{sec:qdof}). In
the following sections we describe the three separate parts of our full
Hamiltonian (\ref{eq:model-Heff}) in more detail.

\subsection{Single junction qubit}\label{sec:qubit}
In this work we consider superconducting qubits with a single Josephson
junction. In a quite general form the Hamiltonian for this type of qubit reads
\begin{equation}\label{eq:model-Hqubit}
 H_{S} = E_C(\hat{n}-n_g)^2+\frac12
E_L(\hat\varphi-\varphi_e)^2-E_J\cos\hat\varphi.
\end{equation}
Here $\hat n$ is the number operator of electrons tunneled through the
junction, $E_C$ is the charging energy, $n_g$ is a tunable charge
offset and $\hat\varphi$ the phase difference between the superconducting
leads. Phase difference and charge number operator are conjugate variables
with the relation $[\hat\varphi,\hat n]=2i$. The inductive energy $E_L$
describes a qubit inside a superconducting loop with applied external flux
$\varphi_e$ (e.g. flux- or phase-qubits). Coherent
Cooper pair tunneling through the junction gives rise to the last term in the
Hamiltonian. $E_J$ is the Josephson energy
which depends on the experimental setup. The Josephson energy is
obtained from second order perturbation theory in the tunnel Hamiltonian
$H_T$. The qubits we consider live in a
parameter regime with $E_J\gg E_C$, where single charge effects are negligible.
In this parameter regime we can neglect effects due to parity\cite{1401.5575}. 
The non-linearity of the Josephson junction plays a crucial role for the
superconducting qubit because it 
allows to truncate the Hilbert space of the effective qubit Hamiltonian to the
two lowest energy levels. The
effective two level Hamiltonian for the qubit (\ref{eq:model-Hqubit}) in
its eigenbasis reads
\begin{equation}\label{eq:model-Hqubiteff}
 H_{S}=\frac{\epsilon_{0}}{2}\sigma_z
\end{equation}
with energy splitting $\epsilon_0$ and the Pauli matrix $\sigma_z$. We will use this
effective Hamiltonian to describe the qubit throughout this work, assuming that
their occurs no transition to higher energies and that the detail of the actual realization
of the qubit is not important for what follows.

\subsection{Quasiparticle degrees of freedom}\label{sec:qdof}
The Hamiltonian $H_{R}$ describes the quasiparticle degrees of freedom in the
superconducting leads. We treat both leads as independent BCS superconductors:
\begin{equation}\label{eq:model-HBCS}
 H_{R}=\sum\limits_{\alpha=l,r} H_{\alpha},\qquad H_\alpha = \sum\limits_k
E_{\alpha,k}^{}\gamma_{\alpha,k\sigma}^\dagger\gamma_{\alpha,k\sigma}^{}
\end{equation}
$\gamma_{\alpha,k\sigma}^{(\dagger)}$ are Bogoliubov
annihilation (creation) operators for quasiparticles with momentum $k$, spin
$\sigma$ and energy$E_{\alpha,k}=(\xi_{\alpha,k}^2+\Delta_\alpha^2)^{1/2}$  in
lead $\alpha$. $\xi_{\alpha,k}$ is the corresponding electron energy in the
normal state measured from the chemical potential.
We assume identical superconductors on either side of the junction
which describes the usual experimental setup. 
Due to the presence of hot non-equilibrium quasiparticles the quasiparticle
distribution functions
\begin{equation}
 f_\alpha(E_k) = \left\langle 
\gamma_{\alpha,k\sigma}^\dagger\gamma_{\alpha,k\sigma}^{} \right\rangle,
\end{equation}
differ from equilibrium Fermi
functions. High energy quasiparticle excitations decay
fast to the gap, e.g. due to inelastic phonon scattering, and produce a
strongly increased quasiparticle density at $E_{\alpha,k}\simeq
\Delta$. Hence, the distribution function differs from a Fermi distribution only in
a
very narrow region above the gap. For temperatures well below the gap the
distribution function decays rapidly for higher energies and $f(E_k)\ll1$ for
$E_k\gtrsim \Delta$. We will assume spin
independent distributions which is in general a good approximation. For some
calculations we describe non-equilibrium quasiparticles with a Fermi or
Boltzmann distribution at an effective temperature $T^*>T$, where $T$ is the base temperature.

Finally we introduce the normalized
density of states
\begin{equation}
 n(\omega) =
\frac{|\omega|}{\sqrt{\omega^2-\Delta^2}}\Theta(\omega^2-\Delta^2)
\end{equation}
of the BCS superconductors and the quasiparticle density per spin defined as
\begin{equation}
 n_{qp} = 2N_0\int_\Delta^\infty dE n(E)f(E).
\end{equation}
Here $N_0=N(0)$ is the normal state density at the Fermi energy.

\subsection{Quasiparticle Tunneling}\label{sec:tunnel}
\paragraph{Electron tunneling}The last term in the effective Hamiltonian
(\ref{eq:model-Heff}) describes
electron tunneling through the junction,
\begin{equation}
 H_T = t\sum\limits_{kq\sigma}e^{i\hat\varphi/2}c_{r,q\sigma}^\dagger
c_{l,k\sigma}+h.c.
\end{equation}
with electron creation/ annihilation operators $c_{\alpha,k\sigma}^{(\dagger)}$.
The commutation relation $[\hat n,e^{i\hat\varphi/2}]=1$ implies the
representation $e^{i\hat\varphi/2} = \sum_n \ket{n+1}\bra{n}$ for the charge
transfer operator $\hat T\equiv e^{i\hat\varphi/2}$. Therefor this operator,
which carries the phase information of the two superconductors, describes the
charge transfer due to tunneling in the qubit's Hilbert space for $E_J\gg E_C$.
Here, we assumed
a constant and real tunneling matrix element $t$ which relates to the
Josephson energy as $E_{J,0} = \pi^2t^2N_0^2\Delta_0$.
The subscripts in the Josephson energy and the gap denote equilibrium
quantities at zero temperature.

\paragraph{Quasiparticles}For use in perturbation theory we need the tunneling
Hamiltonian expressed with free particle operators. For the BCS
superconductors these are Bogoliubov
quasiparticles with the transformation
\begin{equation}
 \begin{pmatrix}
  c_{k\uparrow}\\ c_{-k\downarrow}^\dagger
 \end{pmatrix}
 = 
 \begin{pmatrix}
  u_k & -v_k\\ v_k & u_k
 \end{pmatrix}
 \begin{pmatrix}
  \gamma_{k\uparrow}\\
  \gamma_{k\downarrow}^\dagger
 \end{pmatrix}
\end{equation}
we find the tunneling Hamiltonian 
\begin{align}
  \notag
H_T= H_{qp}+H_p=
\label{eq:model-HT} t\sum\limits_{kq\sigma}(A_{kq}\gamma_{q\sigma,r}
^\dagger\gamma_ { k\sigma , l } ^ {
}+h.c.)\\+t\sum\limits_{kq\sigma}(\sigma
B_{kq}\gamma_{q\sigma,r}^{}\gamma_{k\bar{\sigma},l}^{}+h.c.).
 \end{align}
Here $u_k^2=1-v_k^2=\frac12 (1+\xi_k/E_k)$ are real coefficients while the
phase dependence has already been taken into account with the charge transfer
operator. In the transformed Hamiltonian the first term $H_{qp}$ with coherence
factor 
\begin{equation}
 A_{kk'}=e^{i\varphi/2}u_{k,l}u_{k',r}-e^{-i\varphi/2}v_{k,l}v_{k',r}
\end{equation}
describes single quasiparticle tunneling while the second term $H_p$ refers to
pair tunneling processes with 
\begin{equation}
 B_{kk'}=e^{i\varphi/2}u_{k,l}v_{k',r}+e^{-i\varphi/2}v_{k,l}u_{k',r}.
\end{equation}
The pair
Hamiltonian provides the main contribution to the Josephson term in the qubit
Hamiltonian $\sim E_J\cos\varphi$ but does not contribute to qubit
decoherence as long as relevant energies are small compared to the gap which is
always the case for dephasing. On the other hand $H_{qp}$ describes single
quasiparticles present in the junction region
which undergo incoherent tunneling processes and ultimately induce qubit
decoherence. In
addition to decoherence, quasiparticles lead to corrections of the qubit energies
in two ways\cite{PhysRevB.84.064517}. Virtual transitions between qubit states
lead to a change in energy levels. Second, quasiparticles change physical
parameters of the junction. Both, $E_J$ and $\Delta$, change linearly with the
ratio between quasiparticle and Cooper pair density
$x_{qp}=n_{qp}/(2N_0\Delta_0)$. The resulting corrections to the qubit energy splitting
have been derived by Catelani et all.\cite{PhysRevB.84.064517}. This
work will focus on decoherence effects and we will assume that energy
corrections to the qubit eigenstates have been
included in the effective qubit Hamiltonian.
%
For the qubits considered in this work transitions to higher levels of the
effective qubit Hamiltonian are strongly suppressed due to
the large Josephson energy which minimizes single charge effects.
Hence we truncate the tunneling Hamiltonian to the two dimensional qubit Hilbert
space according to\cite{PhysRevB.72.014517}
\begin{equation}
 \hat T = \alpha I+\vec\beta \cdot \vec\sigma,
\end{equation}
where $I$ is the unit operator in qubit space. We find the coefficients in this
expansion
\begin{align}
 \alpha = \frac12 (\bra{1}\hat T\ket{1}+\bra{0}\hat T\ket{0})\\
 \beta_z = \frac12 (\bra{1}\hat T\ket 1 - \bra 0 \hat T \ket 0)\\
 \beta_x = \bra 1 \hat T \ket 0 + \bra 0 \hat T \ket 1\\
 \beta_y = -i\bra 1 \hat T \ket 0 +i\bra 0 \hat T \ket 1
\end{align}
The terms proportional to $\sigma_{x}$ and $\sigma_y$ induce state transitions.
They describe inelastic tunneling processes with energy exchange between qubit
and bath producing qubit energy relaxation. We will focus on
pure dephasing taking only $\sigma_z$ into account and neglecting other
contributions from quasiparticle tunneling such that $\hat T \rightarrow
\beta_z\sigma_z$ and the tunneling Hamiltonian can be written as
 \begin{align}
  \notag H_T&=\sigma_z
t\sum\limits_{kq\sigma}\left[A^z_{kq}\gamma_{q\sigma,r}^\dagger\gamma_{k\sigma,
l}^{}\right.\\
    \notag&\qquad+\left.\sigma
B^z_{kq}\gamma_{q\sigma,r}^{}\gamma_{k\bar{\sigma},l}^{}+h.c.\right]\\
\label{eq:model-R}
&= \sigma_z (\hat R_{qp}+\hat R_p)\equiv \sigma_z \hat R
 \end{align}
To avoid double counting of processes that have been taken into account in $E_J$
already we have to add an additional term to
$H_T$\cite{PhysRevB.72.014517,PhysRevB.85.144503}:
\begin{equation}
 H_T\rightarrow H_T'=H_T+E_J\cos\varphi.
\end{equation}

\subsection{Spectral density}
The effect of tunneling quasiparticles on
the qubit is described by their spectral density, $S(t) = \langle \hat R(0)\hat
R(t)\rangle$ and its Fourier transform $S(\omega)$. For the quasiparticle part
$H_{qp}$ we find the spectral density
\begin{align}
 \notag
S_{qp}(\omega)=\frac{16}{\pi^2}
\frac{E_J}{\Delta}\int\limits_\Delta^\infty\int\limits_\Delta^\infty
dEdE'\,n(E)n(E')|A(E,E')|^2\\ \label{eq:theory-spectraldensity}\times
f(E)\left(1-f(E')\right)\delta(\omega+E-E')\\
 |A(E,E')|^2 = |\beta_z|^2(1-\cos\vartheta\frac { \Delta^2 } { EE' } )
\label{eq:theory-akq}
\end{align}
with $\cos\vartheta=(\text{Re}[\beta_z]^2-\text{Im}[\beta_z]^2)/(
|\beta_z|^2)$, a interference factor due to the quasiparticle-qubit
interaction \cite{PhysRevB.85.144503,nature13017-4}
This interference plays a crucial role because it determines whether the spectral density diverges or remains finite as $\omega\to0$. If $\cos\vartheta=1$ the singularity due to the BCS density of states cancels out and the spectral density remains finite. For such a qubit the dephasing rate due to quasiparticle tunneling usually remains small and plays only a negligible role. Even for small deviations from $\cos\vartheta=1$ the quasiparticle spectral density is log divergent at zero frequency and it remains open whether this leads to strongly increased dephasing rates.
To conclude this section we have a look at the pair Hamiltonian. It yields the spectral density
\begin{align}
 \notag S_{p}(\omega)=\frac{16}{\pi^2}
\frac{E_J}{\Delta}\int\limits_\Delta^\infty\int\limits_\Delta^\infty
dEdE'\,n(E)n(E')|B(E,E')|^2\\ \notag \times\left\{
\left(1-f(E)\right)\left(1-f(E')\right)\delta(\omega-E-E')\right.\\ \label{eq:theory-bkq}+\left.
f(E)f( E ' ) \delta
(\omega+E+E')\right\}.
\end{align}
For the pair spectral density to become finite we need at least an energy
$|\omega|\gtrsim 2\Delta$. In a usual QED circuit the BCS gap is by far the
largest energy scale, particularly the gap exceeds the relevant energy
scale defined by the energy splitting of the qubit, $\Delta\gg \epsilon_0$.
Therefore we can neglect the pair Hamiltonian from now on and focus on the
 single quasiparticle tunneling described by $H_{qp}$.

\section{Qubit Decoherence}
The effective Hamiltonian (\ref{eq:model-Heff}) describes a small system with
only a few degrees of freedom (qubit) coupled to large reservoirs (leads). 
Tracing out quasiparticle degrees of freedom we find the reduced density matrix
$\rho(t)=\text{Tr}_{qp}\left[\varrho(t)\right]$ of the qubit. We denote the
full density matrix with $\varrho(t)$ and the reduced density matrix, describing only the qubit,
with $\rho(t)$.
Assuming that the full density matrix factorizes at some initial time $t_0$,
$\varrho(t_0)=\rho(t_0)\rho_{qp}(t_0)$, we find an exact relation for the matrix
elements $\rho_{ss'}=\bra{s}\rho\ket{s'}$:
\begin{align}
 \notag\rho_{ss'}(t)=&e^{-i(E_s-E_{s'})(t-t_0)}\sum\limits_{qq'}\rho_{qq'}
(t_0)\\ & \label{eq:theory-rhooft}
\times\text{Tr}_{qp}\left\{\bra{q'}U_I^\dagger(t,t_0)\ket{s'}\bra{s}U_I(t,
t_0)\ket{q}\rho_{R}(t_0)\right\}.
\end{align}
where $\text{Tr}_{qp}\{\dots\}$ denotes a trace with respect to
quasiparticle states, $\ket{s}$ are qubit states and $U_I(t,t')$ is the time
evolution operator in the interaction picture
\begin{equation}
  U_I(t,t_0)=\text{Texp}\left\{-i\int\limits_{t_0}^tH_T(t')dt'\right\}.
\end{equation}
Expanding the time evolution operators in equation (\ref{eq:theory-rhooft}) we
find a real time diagrammatic series for the time
evolution of the the reduced density matrix. With the time evolution superoperator $\Pi(t,t_0)$ we can rewrite the
matrix time evolution as $\rho(t)=\Pi(t-t_0)\rho(t_0)$ where the time evolution
superoperator's matrix elements satisfy the master equation
\begin{align}
 \notag\dot\Pi_{ss'\leftarrow qq'}(t,t_0) = i(E_s-E_{s'})\Pi_{ss'\leftarrow
qq'}(t,t_0)\\\label{eq:theory-master}
+\sum\limits_{q_1q_1'}\int\limits_{t_0}^t\Sigma_{ss'\leftarrow
q_1q_1'}(t,t')\Pi_{q_1q_1'\leftarrow qq'}(t',t_0).
\end{align}
The first term in the Master equation describes free time evolution, while the
kernel $\Sigma(t,t')$ contains all
reservoir effects. In the diagrammatic language we can identify the kernel
with the selfenergy, the sum of all irreducible diagrams. The diagrammatic
approach to the full time evolution is quite general. 
However, for pure dephasing with $H_T =
\sigma_z \hat R$ (\ref{eq:model-R}) the perturbation is diagonal in qubit space and we can simplify the problem. Instead of dealing with the full superoperator $\Pi_{deph}$ we can separate the free time evolution of the qubit states from the noise induced  incoherent time evolution $F(t,t_0)$ as $\Pi_{deph}(t,t_0)=\Pi_0(t-t_0)e^{\Gamma_1 t/2}F(t-t_0)$.The incoherent time evolution is no operator but an ordinary function. 
Later we will show that for quasiparticle tunneling $F(t)\sim e^{-x(t)}$ and will refer to $x(t)$ as dephasing function.
From (\ref{eq:theory-rhooft}), 
it follows
\begin{align}\label{eq:theory-rhodephasing}
  \rho_{ss'}(t)=e^{-i(E_s-E_{s'})(t-t_0)}\rho_{ss'}(t_0)F_{ss'}(t,t_0). 
 \end{align}
 with the incoherent time evolution $F_{ss'}$ defined as
 \begin{align}
  \label{eq:theory-F}F_{ss'}(t,t_0)=
\text{Tr}_{R}\left\{U_I^\dagger(t,t_0,s')U_I(t,t_0,s)\rho_{R}(t_0)\right\}\\
  U_I(t,t_0,s)=T\exp\left\{-is\int\limits_{t_0}^t\hat R(t')dt'\right\}
 \end{align}
where $s=\pm 1$ for excited/ ground state respectively. For $s=s'$
the two time ordered exponentials are inverse to each other so that the diagonal
elements of the dephasing function equal to one. This does not come as a surprise
since the diagonal elements of the density matrix do not feel the longitudinal
coupling and evolve free in time.
Using that the trace over odd powers of the tunneling Hamiltonian vanishes and
demanding a physical density matrix,  $\rho_{01}(t)=\rho_{10}^*(t)$, we find $F_{01}=F_{10}\equiv F(t,t_0)$ where
$F(t,t_0)$ is a real valued function.
 In the following sections we will use a diagrammatic expansion to
calculate $F(t)$. The standard way to obtain dephasing rates uses lowest
order Markov approximation and the corresponding rate is proportional to the
quasiparticle spectral density at zero energy,
\begin{equation}
 \Gamma_{2^*}\sim S_{qp}(0).
\end{equation}
However due to the square root divergent BCS density of states the
quasiparticle spectral density has a logarithmic divergence at zero frequency and the lowest order Markovian dephasing rate is ill defined.
We solve this problem with the introduction of a selfconsistent dephasing
function producing a selfconsistent rate equation, similar
to\cite{PhysRevB.72.014517}.
In the section after we calculate the dephasing function using equation
(\ref{eq:theory-F}) without diagrammatic expansion and Markov approximation.
Pure dephasing is dominated by quasiparticle energies $\omega=E-\Delta\sim 0$. 
Equation (\ref{eq:theory-bkq}) clearly shows, that the pair Hamiltonian contributes to the spectral
density only at energies $\omega\gtrsim2\Delta$. Therefore it does not contribute to pure dephasing and we focus on $H_{qp}$ from now on, neglecting pair contributions.
\subsection{Diagrammatic Expansion - Self consistent rate}
 In this section we use a real time diagrammatic expansion to calculate the
dephasing function $F(t,t_0)$. This expansion is well established in the context
of open quantum systems and we sketch only some steps that are important for our
specific calculation.
 The first step on the way to a diagrammatic description of the problem is to
expand the exponentials in (\ref{eq:theory-F}) and represent the series on a
Keldysh contour. 
The dephasing function is the sum of all diagrams. In order to find the master
equation (\ref{eq:theory-master}) we define the self energy $\Sigma$ in the
usual way as the sum of all irreducible diagrams:
\begin{equation*}\label{eq:diagram-sigmafigure}
  \begin{tikzpicture}[node distance=.75cm and .75cm]
  \matrix[row sep=.5cm,column sep=.05cm] {
   \node[coordinate] (t1) {};
   \node[coordinate,right=of t1] (t2) {};
   \node[coordinate,below=of t1] (b1) {};
   \node[coordinate,right=of b1] (b2) {};
   \path[system] (t1) rectangle (b2);
   \path[] (t1) to node[midway] {\large{$\Sigma$}} (b2);
   &
   \node[coordinate] (ht1) {};
   \node[coordinate] [below=of ht1] (hb1) {};
   \path (ht1) to node[midway] {$=$} (hb1);
   &
   \node[coordinate] (tin) {};
   \node[coordinate,right=of tin] (tout) {};
   \node[coordinate,below=of tin] (bin) {};
   \node[coordinate,right=of bin] (bout) {};
   \node[vertex] (t1) at (tin) {};
   \node[vertex] (b1) at (bout) {};
   \path[system] (t1) -- (tout) (b1) -- (bin);
   \path[bath] (t1) to [bend left=10] (b1);
   \path[bathr] (b1) to [bend left=10] (t1);
   &
   \node[coordinate] (ht1) {};
   \node[coordinate] [below=of ht1] (hb1) {};
   \path (ht1) to node[midway] {$+$} (hb1);
  &
   \node[coordinate] (tin) {};
   \node[coordinate,right=of tin] (tout) {};
   \node[coordinate,below=of tin] (bin) {};
   \node[coordinate,right=of bin] (bout) {};
   \node[vertex] (t1) at (tin) {};
   \node[vertex] (b1) at (bout) {};
   \path[] (tin) to node[pos=0.33,vertex] (t2) {} node[pos=0.66,vertex] (t3) {}
(tout);
   \path[system] (t1) -- (t2) -- (t3) -- (tout) (b1) -- (bin);
   \path[bath] (t1) to [bend right=70] (t3);
   \path[bath] (t2) to [bend left=10] (b1);
   \path[bathr] (t3) to [bend left=50] (t1);
   \path[bathr] (b1) to [bend left=10] (t2);
     
   &
   \node[coordinate] (ht1) {};
   \node[coordinate] [below=of ht1] (hb1) {};
   \path (ht1) to node[midway] {$+$} (hb1);
   &
   \node[coordinate] (tin) {};
   \node[coordinate,right=of tin] (tmiddle) {};
   \node[coordinate,right=of tmiddle] (tout) {};
   \node[coordinate,below=of tin] (bin) {};
   \node[coordinate,right=of bin] (bmiddle) {};
   \node[coordinate,right=of bmiddle] (bout) {};
   \path[] (tin) to node[pos=0,vertex] (t1) {} node[pos=0.2,vertex] (t2) {}
node[pos=0.8,vertex] (t3) {} node[pos=1,vertex] (t4) {} (tout);
   \path[] (bin) to node[pos=0.4,vertex] (b1) {} node[pos=0.6,vertex] (b2) {}
(bout);
   \path[bath] (t3) to [bend left=70] (t1);
   \path[bath] (t2) to [bend left =10] (b1);
   \path[bath] (b2) to [bend right=10] (t4);
   \path[bathr] (t1) to [bend right=50] (t3);
   \path[bathr] (b1) to [bend left=10] (t2);
   \path[bathr] (t4) to [bend right=10] (b2);
   \path[system] (t1)--(t2)--(t3)--(t4) (bout)--(b2)--(b1)--(bin);
   &
   \node[coordinate] (ht1) {};
   \node[coordinate] [below=of ht1] (hb1) {};
   \path (ht1) to node[midway] {$+$} (hb1);
   &
   \node[coordinate] (ht1) {};
   \node[coordinate] [below=of ht1] (hb1) {};
   \path (ht1) to node[midway] {$\cdots$} (hb1);
   \\};
  \end{tikzpicture} 
\end{equation*}
Here solid dots represent tunneling vertices, a dashed directed line is a
contraction in the right lead while a directed solid line represents a
contraction in the left lead and horizontal lines are free time evolution which,
for the dephasing function $F$, equals to one. We restrict the series to
diagrams with two vertex fermionic loops. In other words a left lead contraction
between two vertices implies a right lead contraction between the same vertices.
Since we assume identical superconductors the direction within each loop is
unimportant and we can combine the two possible directions into one contraction
between time $t$ and time $t'$. Each of those contractions
$\gamma^{\gtrless}(t-t')$ yields the quasiparticle spectral density
$S(\pm(t-t'))$ where $\gtrless$ for $t\gtrless t'$ with respect to the Keldysh
contour. With the selfenergy we find the Dyson equation for the dephasing
function $F$ which we illustrate diagrammatically
\begin{equation*}\label{eq:diagram-dyson}
  \begin{tikzpicture}[node distance=.75cm and .75cm]
   \matrix[row sep=.5cm,column sep=.05cm] {
   \node[coordinate] (t1) {};
   \node[coordinate,right=of t1] (t2) {};
   \node[coordinate,below=of t1] (b1) {};
   \node[coordinate,right=of b1] (b2) {};
   \path[system] (t1) rectangle (b2);
   \path[] (t1) to node[midway] {\large{$F$}} (b2);
   &
   \node[crossing] (ht1) {};
   \node[crossing] [below=of ht1] (hb1) {};
   \path (ht1) to node[midway] {$=$} (hb1);
   &
   \node[coordinate] (tin) {};
   \node[coordinate,right=of tin] (tout) {};
   \node[coordinate,below=of tin] (bin) {};
   \node[coordinate,right=of bin] (bout) {};
   \path[system] (tin)--(tout) (bout)--(bin);
   &
   \node[crossing] (ht1) {};
   \node[crossing] [below=of ht1] (hb1) {};
   \path (ht1) to node[midway] {$+$} (hb1);
   &
   \node[coordinate] (t1) {};
   \node[coordinate,right=of t1] (t2) {};
   \node[coordinate,below=of t1] (b1) {};
   \node[coordinate,right=of b1] (b2) {};
   \path[system] (t1) rectangle (b2);
   \path[] (t1) to node[midway] {\large{$F$}} (b2);
   \node[coordinate,right=of t2] (t3) {};
   \node[coordinate,right=of b2] (b3) {};
   \path[system] (t2) rectangle (b3);
   \path[] (t2) to node[midway] {\large{$\varSigma$}} (b3);
   \node[coordinate,right=of t3] (t4) {};
   \node[coordinate,right=of b3] (b4) {};
   \path[system] (t3) -- (t4) (b4)--(b3);
   \\};
  \end{tikzpicture}
\end{equation*}
The time derivate of the Dyson equation shown above yields the master equation
(\ref{eq:theory-F}) for the dephasing function. Assuming a memoryless bath we
can apply Markov approximation. For a memoryless bath it holds that the kernel
$\Sigma(t)$ decays on time scales much shorter then typical qubit decay times.
In this case the integration region in (\ref{eq:theory-master}) is effectively
reduced to a narrow region around $t=t'$. The dephasing function $F(t')$ is
constant in this region and can be replaced by its value at time $t$. In this
approximation quasiparticle tunneling produces an exponential decay with rate
\begin{equation}\label{eq:diagram-markovrate}
 \Gamma_{2^*} = \lim\limits_{\eta\rightarrow0}\int\limits_{-\infty}^0 dt\,
\varSigma(t)e^{\eta t}
\end{equation}
where $\eta$ ensures convergence. In first order four diagrams
contribute to $\varSigma(t)$:
\begin{center}
 \begin{tikzpicture}[node distance=.75cm and .75cm]
 \matrix[row sep=.5cm,column sep=.05cm] {
 \node[coordinate] (t1) {};
   \node[coordinate,right=of t1] (t2) {};
   \node[coordinate,below=of t1] (b1) {};
   \node[coordinate,right=of b1] (b2) {};
   \path[system] (t1) rectangle (b2);
   \path[] (t1) to node[midway] {\large{$\varSigma^{(1)}$}} (b2);
   &
   \node[coordinate] (ht1) {};
   \node[coordinate] [below=of ht1] (hb1) {};
   \path (ht1) to node[midway] {$=$} (hb1);
   &
   \node[coordinate] (tin) {};
   \node[coordinate,right=of tin] (tout) {};
   \node[coordinate,below=of tin] (bin) {};
   \node[coordinate,right=of bin] (bout) {};
   \node[vertex] (t1) at (tin) {};
   \node[vertex] (b1) at (bout) {};
   \path[system] (t1) -- (tout) (b1) -- (bin);
   \path[bathNA] (t1) to [bend left=10] (b1);
   \path[bathrNA] (b1) to [bend left=10] (t1);
   &
   \node[coordinate] (ht1) {};
   \node[coordinate] [below=of ht1] (hb1) {};
   \path (ht1) to node[midway] {$+$} (hb1);
   &
   \node[coordinate] (tin) {};
   \node[coordinate,right=of tin] (tout) {};
   \node[coordinate,below=of tin] (bin) {};
   \node[coordinate,right=of bin] (bout) {};
   \node[vertex] (t1) at (tin) {};
   \node[vertex] (t2) at (tout) {};
   \path[system] (t1) -- (t2) (bin) -- (bout);
   \path[bathNA] (t1) to [bend right=70] (t2);
   \path[bathrNA] (t2) to [bend left=50] (t1);
   &
   \node[coordinate] (ht1) {};
   \node[coordinate] [below=of ht1] (hb1) {};
   \path (ht1) to node[midway] {$+$} (hb1);
   &
   \node[coordinate] (tin) {};
   \node[coordinate,right=of tin] (tout) {};
   \node[coordinate,below=of tin] (bin) {};
   \node[coordinate,right=of bin] (bout) {};
   \node[vertex] (t1) at (tout) {};
   \node[vertex] (b1) at (bin) {};
   \path[system] (t1) -- (tin) (b1) -- (bout);
   \path[bathNA] (t1) to [bend left=10] (b1);
   \path[bathrNA] (b1) to [bend left=10] (t1);
   &
   \node[coordinate] (ht1) {};
   \node[coordinate] [below=of ht1] (hb1) {};
   \path (ht1) to node[midway] {$+$} (hb1);
   &
   \node[coordinate] (tin) {};
   \node[coordinate,right=of tin] (tout) {};
   \node[coordinate,below=of tin] (bin) {};
   \node[coordinate,right=of bin] (bout) {};
   \node[vertex] (t1) at (bin) {};
   \node[vertex] (t2) at (bout) {};
   \path[system] (t1) -- (t2) (tin) -- (tout);
   \path[bathNA] (t1) to [bend left=70] (t2);
   \path[bathrNA] (t2) to [bend right=50] (t1);
   \\};
 \end{tikzpicture}
\end{center}
The first two diagrams each yield $S(t)$ while each of the remaining two yields
$S(-t)$. With the Fourier transformed spectral density we find
\begin{align}
  \notag\Gamma_{2^*} = -2\lim\limits_{\eta\rightarrow
0} \int d\omega S_{qp}(\omega)\int\limits_0^\infty dt\,e^{-\eta
t}\cos\omega t\\
 = 2\pi\lim\limits_{\eta\rightarrow
0}
\int d\omega S_{qp}(\omega)\frac{1}{\pi}\frac{\eta}{\omega^2+\eta^2
}.
 \end{align}
In the given limit the Lorentzian in the latter equation yields a delta function and
we find the dephasing rate $J_{2^*}\sim S_{qp}(0)$. Unfortunately the spectral
density defined in (\ref{eq:theory-spectraldensity}) has a logarithmic
divergence for $\omega=0$. Thus the first order Markovian rate is ill defined
and we need to reconsider our calculations. We want to notice that there exists
an exception to this statement. A closer look on the spectral density reveals
that the divergence is canceled for $\cos\vartheta=1$ which is the case for an
symmetric Hamiltonian. A qubit to which this applies in good approximation is the transmon ($E_L = 0$, $E_J\gg E_C$).
 In this case one finds
\begin{equation}\label{eq:dominant_transmon}
 \Gamma_{2^*} = \frac{32|\beta_z|^2}{\pi}\frac{E_J}{\Delta}f(\Delta).
\end{equation}
Even for the transmon the Hamiltonian $H\sim(n-n_g)^2$ is not strictly symmetric due to the gate charge/ offset charge $n_g$. However since $E_C\ll E_J$ the influence of $n_g$ is exponentially small and as we will show later one can regularize the log divergence in the rate. The regularized rate is not large enough to counter the exponentially small matrix element due to the almost symmetric Hamiltonian and (\ref{eq:dominant_transmon}) remains the dominating contribution to the transmon dephasing rate. 

\subsection{Selfconsistent Born-Markov}
Between vertices in the first order selfenergy a free time evolution occurs
which induces no native decay into the selfenergy. 
However, it is possible 
to generate convergence by including the decay of the propagator $F(t,t')$.
This is achieved within a
selfconsistent Born approximation for the selfenergy and the dephasing function
$F$. We replace the free propagators in the first oder
diagrams with full propagators:
 
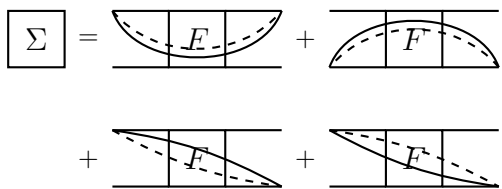
\begin{center}
 \begin{tikzpicture}[node distance=.75cm and .75cm]
   \matrix[row sep=.5cm,column sep=.05cm] {
   \node[coordinate] (t1) {};
   \node[coordinate,right=of t1] (t2) {};
   \node[coordinate,below=of t1] (b1) {};
   \node[coordinate,right=of b1] (b2) {};
   \path[system] (t1) rectangle (b2);
   \path[] (t1) to node[midway] {\large{$\Sigma$}} (b2);
   &
   \node[crossing] (ht1) {};
   \node[crossing] [below=of ht1] (hb1) {};
   \path (ht1) to node[midway] {$=$} (hb1);
   &
   \node[coordinate] (t1) {};
   \node[coordinate,right=of t1] (t2) {};
   \node[coordinate,below=of t1] (b1) {};
   \node[coordinate,right=of b1] (b2) {};
   \path[system] (t1)--(t2) (b2)--(b1);
   \node[coordinate,right=of t2] (t3) {};
   \node[coordinate,right=of b2] (b3) {};
   \path[system] (t2) rectangle (b3);
   \path[] (t2) to node[midway] (s1) {\large{$F$}} (b3); 
   \node[coordinate,right=of t3] (t4) {};
   \node[coordinate,right=of b3] (b4) {};
   \path[system] (t3)--(t4) (b3)--(b4);
   \path[bathNA] (t1) [bend right=70] to (t4);
   \path[bathrNA] (t1) [bend right=50] to (t4);
   &
   \node[crossing] (ht1) {};
   \node[crossing] [below=of ht1] (hb1) {};
   \path (ht1) to node[midway] {$+$} (hb1);
   &
   \node[coordinate] (t1) {};
   \node[coordinate,right=of t1] (t2) {};
   \node[coordinate,below=of t1] (b1) {};
   \node[coordinate,right=of b1] (b2) {};
   \path[system] (t1)--(t2) (b2)--(b1);
   \node[coordinate,right=of t2] (t3) {};
   \node[coordinate,right=of b2] (b3) {};
   \path[system] (t2) rectangle (b3);
   \path[] (t2) to node[midway] (s1) {\large{$F$}} (b3); 
   \node[coordinate,right=of t3] (t4) {};
   \node[coordinate,right=of b3] (b4) {};
   \path[system] (t3)--(t4) (b3)--(b4);
   \path[bathNA] (b4) [bend right=70] to (b1);
   \path[bathrNA] (b4) [bend right=50] to (b1);
   \\
   &
   \node[crossing] (ht1) {};
   \node[crossing] [below=of ht1] (hb1) {};
   \path (ht1) to node[midway] {  $+$  } (hb1);
   &
   \node[coordinate] (t1) {};
   \node[coordinate,right=of t1] (t2) {};
   \node[coordinate,below=of t1] (b1) {};

   \node[coordinate,right=of b1] (b2) {};
   \path[system] (t1)--(t2) (b2)--(b1);
   \node[coordinate,right=of t2] (t3) {};
   \node[coordinate,right=of b2] (b3) {};
   \path[system] (t2) rectangle (b3);
   \path[] (t2) to node[midway] (s1) {\large{$F$}} (b3); 
   \node[coordinate,right=of t3] (t4) {};
   \node[coordinate,right=of b3] (b4) {};
   \path[system] (t3)--(t4) (b3)--(b4);
   \path[bathNA] (t1) [bend left = 10] to (b4);
   \path[bathrNA] (t1) [bend right = 10] to (b4);
   &
   \node[crossing] (ht1) {};
   \node[crossing] [below=of ht1] (hb1) {};
   \path (ht1) to node[midway] {$+$} (hb1);
   &
   \node[coordinate] (t1) {};
   \node[coordinate,right=of t1] (t2) {};
   \node[coordinate,below=of t1] (b1) {};
   \node[coordinate,right=of b1] (b2) {};
   \path[system] (t1)--(t2) (b2)--(b1);
   \node[coordinate,right=of t2] (t3) {};
   \node[coordinate,right=of b2] (b3) {};
   \path[system] (t2) rectangle (b3);
   \path[] (t2) to node[midway] (s1) {\large{$F$}} (b3); 
   \node[coordinate,right=of t3] (t4) {};
   \node[coordinate,right=of b3] (b4) {};
   \path[system] (t3)--(t4) (b3)--(b4);
   \path[bathrNA] (t1) [bend left = 10] to (b4);
   \path[bathNA] (t1) [bend right = 10] to (b4);
   \\};
 \end{tikzpicture}
\end{center}
Within this approximation we are able to sum up all diagrams that
belong to a subclass we call 'boxed'. In this context 'boxed' refers to all
diagrams where the earliest and latest vertex (with respect to real time) are
contracted, the second and second last are contracted and so forth. Some
diagrams of the boxed type:
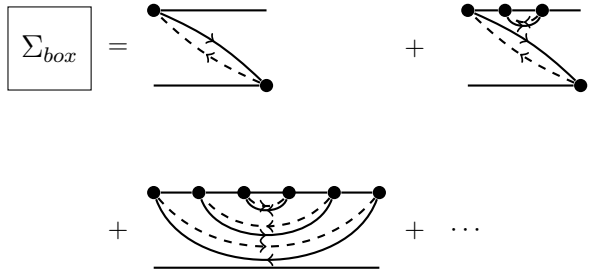
\begin{center}
\begin{tikzpicture}[node distance=1cm and 1.5cm]
  \matrix[row sep=1cm, column sep=.1cm] {
   \node [crossing] (ht1) {};
   \node [crossing,below=of ht1] (hb1) {};
   \path[] (ht1) to node[midway,rectangle,minimum size=1cm+3pt,draw]
{\large{$\Sigma_{box}$}} (hb1);
   &
   \node[coordinate] (ht1) {};
   \node[coordinate] [below=of ht1] (hb1) {};
   \path (ht1) to node[midway] {$=$} (hb1);
   &
   \node[coordinate] (tin) {};
   \node[coordinate,right=of tin] (tout) {};
   \node[coordinate,below=of tin] (bin) {};
   \node[coordinate,right=of bin] (bout) {};
   \node[vertex] (t1) at (tin) {};
   \node[vertex] (b1) at (bout) {};
   \path[system] (t1) -- (tout) (b1) -- (bin);
   \path[bath] (t1) to [bend left=10] (b1);
   \path[bathr] (b1) to [bend left=10] (t1);
   &
   \node[coordinate] (ht1) {};
   \node[coordinate] [below=of ht1] (hb1) {};
   \path (ht1) to node[midway] {$+$} (hb1);
  &
   \node[coordinate] (tin) {};
   \node[coordinate,right=of tin] (tout) {};
   \node[coordinate,below=of tin] (bin) {};
   \node[coordinate,right=of bin] (bout) {};
   \node[vertex] (t1) at (tin) {};
   \node[vertex] (b1) at (bout) {};
   \path[] (tin) to node[pos=0.33,vertex] (t2) {} node[pos=0.66,vertex] (t3) {}
(tout);
   \path[system] (t1) -- (t2) -- (t3) -- (tout) (b1) -- (bin);
   \path[bath] (t2) to [bend right=70] (t3);
   \path[bath] (t1) to [bend left=10] (b1);
   \path[bathr] (t3) to [bend left=50] (t2);
   \path[bathr] (b1) to [bend left=10] (t1);
   \\
   &
   \node[coordinate] (ht1) {};
   \node[coordinate] [below=of ht1] (hb1) {};
   \path (ht1) to node[midway] {$+$} (hb1);
   &
   \node[coordinate] (tin) {};
   \node[coordinate,right=of tin] (tmiddle) {};
   \node[coordinate,right=of tmiddle] (tout) {};
   \node[coordinate,below=of tin] (bin) {};
   \node[coordinate,right=of bin] (bmiddle) {};
   \node[coordinate,right=of bmiddle] (bout) {};
   \path[] (tin) to node[pos=0,vertex] (t1) {} node[pos=0.2,vertex] (t2) {}
node[pos=0.8,vertex] (t3) {} node[pos=1,vertex] (t4) {} (tout);
   \path[] (tin) to node[pos=0.4,vertex] (b1) {} node[pos=0.6,vertex] (b2) {}
(tout);
   \path[bath] (t4) to [bend left=70] (t1);
   \path[bath] (t2) to [bend right =70] (t3);
   \path[bath] (b2) to [bend left=70] (b1);
   \path[bathr] (t1) to [bend right=50] (t4);
   \path[bathr] (t3) to [bend left=50] (t2);
   \path[bathr] (b1) to [bend right=50] (b2);
   \path[system] (t1)--(t2)--(t3)--(t4) (bout)--(bin);
   &
   \node[coordinate] (ht1) {};
   \node[coordinate] [below=of ht1] (hb1) {};
   \path (ht1) to node[midway] {$+$} (hb1);
   &
   \node[coordinate] (ht1) {};
   \node[coordinate] [below=of ht1] (hb1) {};
   \path (ht1) to node[midway] {$\cdots$} (hb1);
   \\};
  \end{tikzpicture}
\end{center}
We find the self-consistent selfenergy
\begin{equation}
 \Sigma(t) = 2(S_{qp}(t)+S_{qp}(-t))F(t).
\end{equation}
In Markov approximation we know the solution of the master equation
(\ref{eq:theory-master}) for the dephasing function is a simple exponential
decay with rate $\Gamma_{2^*}$, $F(t)=\exp(-\Gamma_{2^*}t)$. With this ansatz
for $F(t)$ and the definition (\ref{eq:diagram-markovrate}) for the Markovian
dephasing rate we find the selfconsistent dephasing rate
\begin{align}\label{eq:sc_gamma}
  \notag\Gamma_{2^*} = -2\int d\omega S_{qp}(\omega)\int\limits_0^\infty
dt\,e^{-\Gamma_{2^*}
t}\cos\omega t\\
 = 2\pi
\int d\omega
S_{qp}(\omega)\frac{1}{\pi}\frac{\Gamma_{2^*}}{\omega^2+\Gamma_{2^*}^2
}.
\end{align}

\subsection{Beyond Markov - Tunneling as  bosonic noise}
In this section we sum up all diagrams which include only two-vertex fermionic
loops, the class of diagrams we have been using throughout this paper. We start
from the definition (\ref{eq:theory-F}) of the incoherent time evolution $F(t)$ and introduce the contour time
ordering $T_C$ which orders operators with respect to the Keldysh contour
introduced in the previous section,
 \begin{equation} 
F(t-t_0)=\text{Tr}_{qp}\left\{T_C\exp\left\{i\int\limits_{t_0}
^t\hat R(t')dt'\right
 \}\right\}.
 \end{equation}
with $H_{qp} = \sigma_z \hat R$. The quasiparticle operator $\hat R$ is defined
as the single particle part in (\ref{eq:model-R}). We notice that this operator
is bilinear in fermionic operators (though it is linear for each lead
separately). In the diagrams we include in our approximation only correlations
between full $\hat R$ operators occur. Hence the bath behaves similar to a bath
linear in bosonic operators. Therefore we expect to find the same behavior as
for a bosonic bath. To confirm this suspicion we expand the
exponentials and use Wick's theorem to calculate the trace over reservoir
states:
 \begin{align}
  \notag F(t)&=\sum\limits_{n=0}^\infty \frac{(-1)^n}{(2n)!}\idotsint
dt_1\cdots dt_{2n}\langle T_C[\hat R(t_1)\cdots \hat R(t_{2n})]\rangle\\
  &=\sum\limits_{n=0}^\infty
\frac{(-1)^n}{(2n)!}\sum\limits_P\prod\limits_{\{ij\}\in P}\iint dt_idt_j\langle
T_C[\hat R(t_i)\hat R(t_j)]\rangle.
 \end{align}
Here $P$ denotes all permutations of time arguments in the trace. Exchanging
two operators does not produce a minus sign in this case since the bath operator
$R$ is bilinear in fermionic operators. The contour ordered bath correlation
function $\langle T_C[R(t_i) R(t_j)]\rangle$ is just the first order self
energy and we find
 \begin{equation}
  F(t) = \exp\left[-2\int\limits_{-\infty}^\infty
d\omega S_{qp}(\omega)\int\limits_{t_0}^tdt_1\int\limits_{t_0}^{t_1}dt_2
\cos(\omega(t_1-t_2))\right]
 \end{equation}
This yields an exponential decay with nonlinear and time dependent dephasing
'rate', $x_2(t)$, where $x_2(t)$ is the known dephasing
function for a Ramsey experiment with bosonic noise\cite{PhysRevB.72.134519}:
 \begin{equation}\label{eq:Ramsey}
x_{2}(t)=t^2\int d\omega S_{qp}(\omega)\text{sinc}^2\left(\frac{\omega
t}{2}\right)
 \end{equation}
Experiments suggest that dephasing times due to quasiparticle tunneling are at
least in the order of micro seconds while typical quasiparticle energies are of
order $\Delta\sim100$GHz. The quadratic sinc function suppresses the
integrand
at values $\omega t\sim\mathcal O(1)$. So the spectral density is
evaluated at energies $\omega \sim 10^6/s$ which on the scale
of quasiparticle energies implies $\omega\approx 0$. The largest
contribution to the dephasing rate still arises from small frequencies.
Since our treatment of quasiparticle tunneling is equivalent to a bosonic bath we can not only explain a Ramsey experiment, which describes off-diagonal element decay after initial preparation, with decay function
(\ref{eq:Ramsey}) but any measurement protocol with different pulse
sequences\cite{nphys1994}.
\section{Results}
\subsection{Analytical results}
In this section we calculate both selfconsistent- and non Markovian dephasing for a narrow quasi particle distribution above the
gap, $f(E)\approx 0$ for $E\gtrsim\Delta$ such that the relation 
\begin{equation}\label{eq:nqp-integral}
 S_{qp}(\omega)=\int\limits_1^\infty J(\omega/\Delta,x)n(x)f(x)=x_{qp}J(\omega/\Delta,1)
\end{equation}
holds where $x_{qp}$ is the quasiparticle density normalized to Cooper pair
density,
\begin{equation}
 x_{qp} = \frac{1}{2\Delta N_0}n_{qp} = \int\limits_1^\infty f(x)n(x)
\end{equation}
and $x = E/\Delta$ is the normalized quasiparticle energy. This kind of quasiparticle distribution reflects the experimental situation quite 
accurately. While quasiparticles may be generated even at higher energies they decay rapidly to
the gap due to inelastic phonon scattering while quasiparticle recombination is rather slow compared to relaxation. 
This kind of processes lead to a pronounced density of quasiparticles with energy close to the gap while for higher energies the distribution is thermal.
For this kind of system we find the quasiparticle spectral density 
\begin{align}
 S_{qp}(\omega) =
\frac{16E_j}{\pi^2}x_{qp}\frac{1+\omega-\cos\vartheta}{\sqrt{2\omega}}.
\end{align}
This yields the selfconsistent rate equation for the normalized dephasing rate $\gamma\equiv\Gamma_{2^*}/\Delta$
\begin{equation}\label{eq:sc_gamma_appr}
 \gamma =
\frac{32E_j}{\pi\Delta}x_{qp}\frac{1+\gamma-\cos\vartheta}{\sqrt{
\gamma}}.
\end{equation}
Experiments suggest that $\Gamma\lesssim1$MHz while $\Delta\sim10^2$GHz and we
expect $\gamma\ll1$. For $\cos\vartheta\neq1$ we find:
\begin{equation}
 \gamma \approx \left(\frac{32E_jx_{qp}}{\pi\Delta}(1-\cos\vartheta)\right)^{2/3}
\end{equation}
while for $\cos\vartheta=1$
\begin{equation}
 \gamma = \left(\frac{32E_jx_{qp}}{\pi\Delta}\right)^2.
\end{equation}
For the approximated spectral density we find the non-Markovian dephasing
function for boson like noise is given by
\begin{equation}\label{eq:xanalytical}
 x_2(t) =
\frac{32E_jx_{qp}}{\pi\Delta}\frac{1}{\sqrt{\pi}}\left[\frac43(1-\cos\vartheta)(\Delta t)^{
3/2}+2(\Delta t)^{1/2}\right].
\end{equation}
For the case $\cos\vartheta=1$ we can define a rate as $x(t)\sim(\gamma_2 \Delta t)^{1/2}$
with 
\begin{equation}\gamma_2 =
\frac{4}{\pi}\left[\frac{32E_jx_{qp}}{\pi\Delta}\right]^2=\frac{4}{\pi}\gamma
\end{equation}
while for $\cos\vartheta\neq1$ we find with $x(t)\sim(\gamma_2 \Delta t)^{3/2}$ a rate
\begin{equation}
\gamma_2=\left(\frac{4}{3\sqrt{\pi}}\right)^{2/3}
\left[\frac{32E_jx_{qp}}{\pi\Delta}(1-\cos\vartheta)\right]^{2/3} = \left(\frac{4}{3\sqrt{\pi}}\right)^{2/3}\gamma.
\end{equation}
We find that the non Markovian 'rates' as defined above are proportional to
the rates obtained from the self-consistent Born approximation but with a different time dependence. It remains the question of
applicability of this approximation. The basic idea behind it is a very narrow
distribution function $f(x)$ with width $\Lambda=\delta E/\Delta\ll1$. To apply
the integral approximation (\ref{eq:nqp-integral}) the function $S_{qp}(x)$ should
not change much within the region $0\leq x\leq\Lambda$. For the quasiparticle
spectral density this holds for $\omega>\Lambda$ so that the approximation
works quite good for qubit decay where $\omega\sim\epsilon_0$. Dephasing on the
other hand is dominated by frequencies $\omega\sim0$. Especially in the long
time limit the sinc function decays for frequencies in the MHz regime and
$\omega\lesssim\delta E$ reflects this situation more realistically. We can estimate the time scale on which this approximation is valid for a Fermi distribution with $\mu\approx\Delta$. In this case the quasiparticle width $\Lambda$ above the gap is of order temperature, $\Lambda\sim k_BT$. In this case the approximation is valid for a time scale up to ($k_B=\hbar=1$) $t\lesssim T^{-1}$ For a temperature $T=2.3mK=0.001\Delta$ for aluminum we find that the approximation should work up to $t\approx10^{-2}\mu s$. In Fig \ref{subfig:xe_over_t_5b} we can approximately confirm this result for $\cos\vartheta\neq1$. In general for
this approximation to be valid an combination of relative slow
dephasing rates and large frequencies must be fulfilled. This can hold only for short times in a typical qubit setup. For longer times, e.g. smaller frequencies, we expect that the non Markovian dephasing becomes linear in time and approaches the selfconsistent dephasing rates for $t\to\infty$. 
For short times the approximation reveals an interesting result since the time dependence is - as we expected
- somewhere between a linear decay for a regular spectral density and a $t^2$
decay for $1/f$ noise, in this case $\sim t^{3/2}$. For $\cos\vartheta=1$ the
spectral density tends to zero as $\omega$ approaches zero and dephasing is
even slower than linear decay - $\sim t^{1/2}$ for small times. In the next section we compare the analytical result with numerical calculations of the full non Markovian dephasing. We confirm the expected behavior $x(t)\sim t^{n/2}$ with $n=2\pm1$ for $t\ll 1\mu s$ while for $t \gtrsim\mathcal O(10^{-2}\mu s)$ the dephasing becomes linear. 
\begin{figure*}[htbp]
  \subfigure[][\label{subfig:xe_over_t_5a}]{%
    \includegraphics[width=.5\linewidth]{xe_over_t_Boltzmann}}%
 \subfigure[][\label{subfig:xe_over_t_5b}]{%
    \includegraphics[width=.5\linewidth]{xe_over_t_Fermi}}%
    \caption{\label{fig:xe_over_t}Normalized non Markovian dephasing function over time $(x_2(t)/t)$ versus $t$ for different interference angles $\vartheta$ for \subref{subfig:xe_over_t_5a} an effective Boltzmann distribution with $T=0.1\Delta\approx230mK$ and \subref{subfig:xe_over_t_5b} an effective Fermi distribution with $T=10^{-3}\Delta\approx 2.3mK$ and $\mu=0.992\Delta$. Both distribution functions yield a quasiparticle density $x_{qp}\sim\mathcal{O}\left(10^{-5}\right)$. Horizontal dashed lines are the corresponding selfconsistent rates $\gamma$ while dot-dashed lines are obtained with the analytical approximation (\ref{eq:xanalytical}). For the narrow Fermi distribution the numerical calculation shows a smooth crossover between the analytical approximation and the constant dephasing rate while for the smeared out Boltzmann distribution the analytical approximation does not describe the actual behavior as good as in the first case.}%
\end{figure*}
\subsection{Numerical results}

In this section we calculate the dephasing function and selfconsistent rate numerically. For that we assume two different forms for the
distribution function. Starting from an effective Fermi distribution $f(x) =
1/(1+\exp(\beta(x-\mu))$ we first assume a small chemical potential and a
large effective temperature such that $\beta=\hbar\Delta/k_BT\sim \mathcal
O(1-10)$ and $\beta\mu\ll 1$. In this case we can approximate the Fermi distribution with a
Boltzmann distribution, $f(x)\sim f_0 \exp(-\beta x)$. This distribution
function, due to the large temperature, is smeared out over a wide range of energies above the gap and the
former approximation for a narrow distribution function is likely to fail as it gives to much weight to
quasiparticles at high energies which are effectively suppressed due to the sinc-function or the Lorentzian respectively.
The second case we investigate is that of quasiparticles with a chemical potential close to the
gap such that $\mu=1+\Lambda$ with $\Lambda\ll1$ while the effective temperature remains quite low 
such that the quasiparticle distribution decays in a narrow region above the gap. For such a distribution
function we expect the approximation from previous section to be quite accurate as long as the quasiparticle width $\Lambda$ remains small. 
From the analytical $x_{qp}$-approximation we expect that the dephasing function has a time
dependence according to $at^{3/2}+bt^{1/2}$ for short times while for long times the non Markovian dephasing $x(t)$ should become linear and approach the selfconsistent rates.
All numerical results are obtained for an aluminum transmon with $\Delta = 300\text{GHz} = 200\mu eV$, $E_C = 2\times2\pi$GHz and $E_J=20E_C$. These parameters are used to calculate $|\beta_z|^2$ and the spectral density $S_{qp}$ while we choose arbitrary interference angles $\vartheta$ to demonstrate its influence on dephasing times.
We plot our results in figure \ref{fig:xe_over_t} for a Boltzmann and narrow Fermi distribution respectively. The rates show the
expected time behavior. While the dephasing function
$x_2(t)/t$ decays for the special case $\cos\vartheta=1$ where we
expect $x_2(t)/t\sim t^{-1/2}$ it increases in time for every other value of the
interference factor. Indeed the $x_{qp}$ approximation is valid for narrow distribution function and short times while for longer times we find $x(t)/t \sim \Gamma_{2^*}$ (although it keeps ascending) in good agreement with our expectations.
\subsection{Selfconsistent rate - Dependence on interference angle}
\begin{figure}[htbp]
  \includegraphics[width=\linewidth]{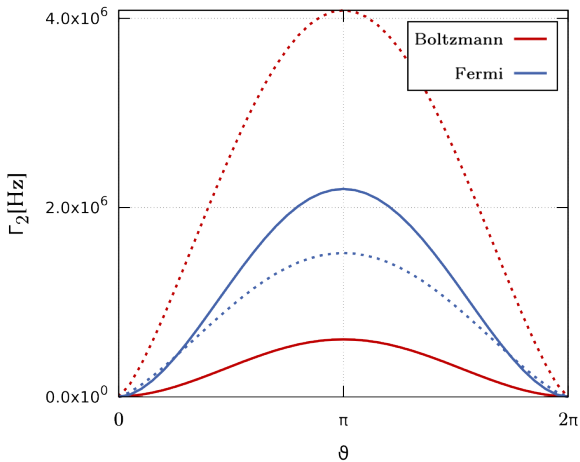}
 \caption{\label{fig:sc_gamma}Selfconsistent rate
$\Gamma_{2^*}$ versus $\vartheta$ for an effective Boltzmann
(blue) and Fermi (red) distribution function. Solid lines are numerical
iterations of the selfconsistent equation (\ref{eq:sc_gamma}) while dashed lines are obtained from the analytical approximation 
(\ref{eq:sc_gamma_appr}). The approximation overestimates the rate strongly for the Boltzmann case. This can readily be explained with due to the small rates $\Gamma\ll1$MHz. The Lorentzian weighting function in the selfconsistent equation has a very narrow peak which measures the spectral density only at small frequencies while for the $x_{qp}$ approximation all quasiparticles are taken into account within the quasiparticle density. For the narrow Fermi distribution the approximation is better because their are no quasiparticles present at higher energies}  
\end{figure}
The effect of tunneling quasiparticles on qubit dephasing is determined by two
main factors: the form of the quasiparticle distribution function and the
qubit matrix element interference factor $\cos\vartheta =(
\text{Re}(\beta_z^2)-\text{Im}(\beta_z^2))/|\beta_z|^2$. For purely real matrix element
$\beta_z$ this factor equals one and the spectral density
tends to zero as $\omega$ approaches zero. In this special case dephasing due
to quasiparticles remains small and can be neglected compared to the $T_1$
time. Any other case offers the possibility of large pure dephasing due to the
log divergence. In figure \ref{fig:sc_gamma} we show the selfconsistent dephasing rate for the same qubit parameters and quasiparticle distribution functions as in the previous section. The rate increases with interference $\vartheta$ until it reaches a maximum at $\vartheta = \pi$. Nonetheless for the Boltzmann distribution it remains small compared to $\Gamma_1$: $\Gamma_{2^*}\sim\mathcal O(0.1\text{MHz})$ while $\Gamma_1 \sim 1.4\text{MHz}$ for the parameters used for the dephasing calculations. On the other hand for the narrow distribution it becomes comparable and must be taken into account for decoherence if $\cos\vartheta\neq 1$. 

\section{Conclusion}
In this work we analyzed qubit dephasing due to quasiparticle tunneling. We applied two different techniques 
to the problem at hand. First we assumed Markovian behavior and used a self-consistent Born
approximation to find a selfconsistent equation for the pure dephasing rate. The
rate defined in equation (\ref{eq:sc_gamma_appr}) is identical to the one defined
in equation (28) from\cite{1207.7084} and we want to recommend this work for a more detailed analysis of the selfconsistent rate. 
But as is known from 1/f noise pure dephasing for a noise with irregular spectral
density at zero frequency does usually not follow a simple exponential law with linear exponent.
We therefore extended our calculations for non-Markovian behavior and found a dephasing function
similar to the Ramsey dephasing functions known for bosonic noise\cite{PhysRevB.72.134519}. 
We showed that pure dephasing obeys exponential decay with exponent $\sim t^{\alpha}$ 
with exponent $\alpha\sim-1/2 \dots 3/2$ for short times while for longer time scales the dephasing function becomes more and more linear and similar to the selfconsistently obtained rate. The pure dephasing
remains small and does not limit qubit coherence for the transmon and similar qubits.

\nocite{*}
\bibliography{main}
\addcontentsline{toc}{chapter}{\bibname}
\end{document}